\documentclass[review]{elsarticle}

\usepackage{lineno,hyperref}
\modulolinenumbers[5]
\usepackage{adjustbox}
\usepackage{lipsum,amsmath,multicol}
\usepackage{algorithm}
\usepackage{algorithmic}
\usepackage{subfigure}
\usepackage{array, booktabs, makecell}
\usepackage{graphicx}
\usepackage{caption}
\usepackage{booktabs,multirow}
\usepackage{hhline}
\usepackage{mathtools}
\usepackage{array}
\usepackage{float}
\begin{document}

\captionsetup[figure]{labelfont={bf},labelformat={default},labelsep=period,name={Fig.}}
\begin{frontmatter}

\title{Development of algorithm to model dispersed gas-liquid flow using lattice Boltzmann method}
\author[mymainaddress]{Alankar Agarwal\corref{mycorrespondingauthor}}
\cortext[mycorrespondingauthor]{Corresponding author}
\ead{agarwal.1@iitj.ac.in}
\author[mymainaddress]{B. Ravindra}
\author[mysecondaryaddress]{Akshay Prakash}

\address[mymainaddress]{Indian Institute of Technology Jodhpur, Jodhpur, Rajasthan, India-342037}
\address[mysecondaryaddress]{Indian Institute of Technology Kharagpur, Kharagpur, West Bengal, India-721302}

\begin{abstract}
In this paper, we present the algorithm for the simulation of a single bubble rising in a stagnant liquid using Euler-Lagrangian (EL) approach. The continuous liquid phase is modeled using BGK approximation of lattice Boltzmann method (LBM), and a Lagrangian particle tracking (LPT) approach has been used to model the dispersed gas (bubble) phase. A two-way coupling scheme is implemented for the interface interaction between two phases. The simulation results are compared with the theoretical and experimental data reported in the literature and it was found that the presented modeling technique is in good agreement with the theoretical and experimental data for the relative and terminal velocity of
a bubble. We also performed the grid independence test for the current model and the results show that the grid size does not affect the rationality of the results. The stability test has been done by finding the relative velocity of a bubble as a function of time for the different value of dimensionless relaxation frequency. The present study is relevant for understanding the bubble-fluid interaction module and helps to develop the accurate numerical model for bioreactor simulation.
\end{abstract}


\end{frontmatter}

\section{Introduction}
Bubble column reactors are widely encountered in chemical, biological and pharmaceutical industry. The motion of dispersed air bubble in its process has been the focus of research for a long time \cite{frank06}. In the past few decades, a number of experimental investigations \cite{cli78, sadhal97, wen10} have been performed to understand the basic underlying physics and its hydrodynamics \cite{nie15}. Based on the empirical relations obtained from experiments, researchers developed various numerical models or computational fluid dynamics (CFD) techniques for this multiphase problem. These models are categorized based on their treatment of dispersed (gas) phase into continuous (liquid) phase \cite{sung12} as follows:
\begin{itemize}
\item \textbf{Euler-Euler} (\textbf{EE}) model also referred as two-fluid model, where both the dispersed (gas bubble) and continuous (liquid) phase are treated as interpenetrating continua, and interaction between the two phases are modeled using the phase interaction terms that appear in the conservation equations, describes the dynamics of the system \cite{law06},
\item \textbf{Euler-Lagrangian} (\textbf{EL}) model, in which the liquid phase is modeled in eulerian cell, while the dispersed gas bubble is treated as Lagrangian marker. The motion of bubbles is governed by the Newton's law of motion. The Euler-Lagrangian approach requires closure relations for the forces between two-phases, which can be obtained from the experimental correlations or from simulations with higher level of details (e.g volume-of-fluid (VOF) or front-tracking (FT) method) \cite{sung11}.
\end{itemize}

This work is focused on the Eulerian-Lagrangian (EL) approach to simulate the dispersed gas-liquid flow problem. Although a version of this approach has already been reported in the literature, the present study uses the BGK scheme of lattice Boltzmann method (LBM) to model the continuous liquid phase in a Eulerian frame of reference and the motion of dispersed gas bubble is computed using Lagrangian particle tracking (LPT) approach. The two-way coupling for momentum transfer between the phases is achieved with the cheap-clipped polynomial mapping function proposed by \cite{deen04}.

In recent decades, lattice Boltzmann Method (LBM) has emerged as a powerful numerical tool to simulate multiphase flows. The method based on the molecular kinetic theory shows numerous advantages over conventional computational fluid dynamics (CFD) method. The method has been proven to be an efficient algorithm for the simulation of complex boundary problems, and due to explicit nature, it is easily parallelizable \cite{gupta07}. The interested reader is referred to \cite{aid10, chen98, huang14, suc01} for more information about the various multiphase models of lattice Boltzmann method (LBM). 

This paper is organized in the following manner: The methodology to simulate dispersed gas-liquid flow is reported in section 2, which describes the governing equations for bubble and liquid phase hydrodynamics in its subsequent subsection. In section 3, the test flow problem is described along with geometry and simulation parameters, the coupling between two phases and boundary-conditions. The results are discussed in section 4.  Section 5 provides the summary of the work and concludes the paper.
\section{Methodology}

In this study, an air bubble is released in a 3D rectangular column tank filled with stagnant water. The transient, three-dimensional Euler-Lagrange model is used to simulate this multiphase problem. The model consists of two processes: the first process includes the bubble motion and the second part describes the liquid velocity fluctuations. The EL model requires constitutive equations to couple two processes through the forces acting between a bubble and liquid \cite{darm06}. The interaction between the dispersed (gas) phase and continuous (liquid) phase can be modeled with the two-way coupling approach.
\subsection{Bubble dynamics}
The air bubble in a stagnant water tank is treated as a point-volume particle with constant mass \cite{sung11}, the motion of bubble is computed from the Newton$'$s second law of motion:
\begin{equation}
m_{b}(d{\mathbf{u_b}}/dt) = \sum \mathbf{F_b}
\end{equation}
where $\mathbf{u_b}$, $m_b$ , and $\mathbf{F_b}$ are the velocity, mass and total force acting on the bubble respectively.
The net force acting on bubble is composed of several external forces i.e. buoyancy
force $\mathbf{F_B}$, stress gradient force $\mathbf{F_S}$, drag force $\mathbf{F_D}$, lift force $\mathbf{F_L}$ and virtual mass force $\mathbf{F_{VM}}$ gives:
\begin{equation}
\mathbf{F_b} = \mathbf{F_B}+\mathbf{F_S}+\mathbf{F_{VM}}+\mathbf{F_D}+\mathbf{F_L}
\end{equation}
An expression to compute these forces is given in Table 1.
\begin{table}[!h]
\caption{Expression for interface forces acting on bubble \cite{delno97}, \cite{sung11}}
\centering
\begin{adjustbox}{width=1\textwidth}
\small
{\begin{tabular}[l]{@{}l l}
\hline\noalign{\smallskip}
  \textbf{Force} & \textbf{Coefficient relation}\\
\noalign{\smallskip}\hline\noalign{\smallskip}
  $\mathbf{F_B}$ = $(\rho_l - \rho_b)V_bg$ & -\\
$\mathbf{F_S}$ = $\rho_lV_bD_t\mathbf{u_l}$& -\\
$\mathbf{F_{VM}}$ =  $-C_{A}\rho_{l}V_{b}(D_{t}\mathbf{u_{b}}-D_{t}\mathbf{u_{l}})$ & $C_A=0.5$\\
$\mathbf{F_D}$ = $\frac{-1}{2}C_{D} \rho_{l}\pi r_{b}^{2}\mid \mathbf{u_{b}}-\mathbf{u_{l}}\mid(\mathbf{u_{b}}-\mathbf{u_{l}})$ & $C_{D} =max[min{\frac{24}{Re}(1+0.015Re^{0.687}, \frac{48}{Re}},\frac{8}{3}\frac{E_{o}}{E_{o}+4}]$ \\
$\mathbf{F_{L}}$= $-C_{L}\rho_{l}V_{b}(\mathbf{u_{b}}-\mathbf{u_{l}})\times\nabla \times \mathbf{u_{l}}$& $C_{L} = \begin{cases}
min[0.288tanh(0.121Re, f(E_{od}))],& E_{od}<4\\
f(E_{od}),&4<E_{od}\leq 10\\
-0.29,& E_{od} >10
\end{cases}$\\
& $E_{od}$=$\frac{E_{o}}{E^{2\backslash3}}, E = \frac{1}{1+0.163E_{o}^{0.757}}$\\
& $f(E_{od})$ = $0.00105E_{od}^{3}-0.0159E_{od}^{2}-0.0204E_{od}$\\ 
\noalign{\smallskip}\hline
\end{tabular}}
\label{1}
\end{adjustbox}
\end{table}

where the Eotvos number, $E_o$ and Reynolds number, $R_e$ can be calculated as follows \cite{darm06}:
\begin{equation}
E_{o} = (\rho_{l}-\rho_{b})gd_{b}/\sigma, \ \ \
R_{e} = \rho_{l}(\mathbf{u_{b}}-\mathbf{u_{l}})/\mu_{l}
\end{equation}
where, $\rho_l$, $\rho_b$, $\mathbf{u_b}$ and $\mathbf{u_l}$ represent the liquid density, air-bubble density, bubble velocity and liquid velocity respectively.
Due to these forces acting on bubble, it start accelerating. The bubble velocity at next time step can be computed using the expression below:
\begin{equation}
\mathbf{u_{b}^{n+1}}=\mathbf{u_{b}^{n}}+ \Big[\Big(\sum \mathbf{F_{b}\Big)}/m_{b}\Big]\Delta t_{b}
\end{equation}
where, $\Delta t_{b}$ is the time step for dispersed air bubble calculation.

\subsection{Liquid phase hydrodynamics: LBGK model}
The continuous fluid in this gas-liquid flow is discritized using nineteen velocity ($D_3Q_{19}$) lattice Bhatnagar-Gross and Krook (LBGK) model on cubic lattice. The model is also popularly known as single-relaxation time (SRT) model. The schematic representation of $D_3Q_{19}$ lattice structure is shown in Fig.1.
\begin{figure}[hb]
\begin{center}
\begin{adjustbox}{width=0.6\textwidth}
\small
\includegraphics[scale=0.4]{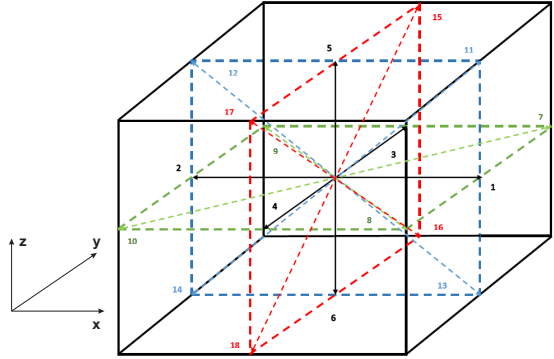} 
\end{adjustbox}
\end{center}
\caption{Shematic representaion of $D_3Q_{19}$ lattice structure}
\label{fig:1}
\end{figure}

The lattice Boltzmann equation (LBE) with single-relaxation time (SRT) parameter without a forcing term can be written as \cite{kang11}:
\begin{equation}
f_{j}(x+c_{j}\Delta t_{l}, t_l+\Delta t_{l}) = f_{j}(x,t_l) - (1/\tau)[f_{j}(x,t_l) - f_{j}^{eq}(x,t_l)]
\end{equation}
where $f$ is the probability density distribution function (PDDF). The equilibrium probability density distribution function can be computed as \cite{bhat54}:
\begin{equation}
f_{j}^{eq} = w_{j}\rho_l\Big[1+ \frac{3}{e^{2}}(c_{j}.\mathbf{u_{l}}) + \frac{9}{2e^{4}}(c_{j}.\mathbf{u_{l}})^{2}-\frac{3}{2e^{2}}\mathbf{u_{l}^{2}}\Big]
\end{equation}
For the  $D_3Q_{19}$ model the discrete velocity vectors $c_j$ , and the corresponding weighted function $w_j$ can be expressed as \cite{perumal2015}:
\begin{equation}
c_{j}=
\begin{cases}
 e(\pm 1, 0, 0), e(0, \pm 1, 0), e(0, 0 \pm 1), & j=1,\dots, 6 \\
 e(\pm 1, \pm 1, 0), e(\pm 1, 0, \pm 1), e(0, \pm 1, \pm 1), & j=7,\dots, 18\\
 e(0,0,0), & j = 19 
\end{cases}
\end{equation}
\begin{equation}
w_{j} = \begin{cases}
1/18,& j=1,\dots, 6 \\
1/36,& j=7,\dots, 18\\
1/3,&  j = 19
\end{cases}
\end{equation}
where the lattice speed $e$ = $\frac{\Delta x}{\Delta t_l}$, and $\Delta x$ and $\Delta t_l$ are the respective lattice size and the time step for the calculation of continuous liquid phase. The dimensionless relaxation time parameter $\tau$ is related to the kinematic viscosity that fixes the rate of approach to equilibrium given by \cite{zhang03}:
\begin{equation}
\nu = \Big((2\tau-1)/6\Big)*\Big((\Delta x)^{2}/\Delta t_l\Big),\ \  \tau = 1/\omega
\end{equation}
where, $\omega$ is the dimensionless relaxation frequency.

The macroscopic variables such as density per node and momentum density are computed from the real-valued PDDF by \cite{bhat54}:
\begin{equation}
\rho_l=  \sum_{j}f_{j}=  \sum_{j}f_{j}^{eq},\ \ \ \  \rho_l\mathbf{u_l}=  \sum_{j}c_{j}f_{j} =  \sum_{j}c_{j}f_{j}^{eq}
\end{equation}
This density and momentum density satisfy the traditional pressure-based solver (i.e. Navier-Stokes solver) for incompressible flow explained by using the Chapman-Enskog expansion \cite{he97}.

When an external force is applied in the computational cell, the LBE equation can be defined as \cite{he197}:
\begin{equation}
f_{j}(x+c_{j}\Delta t_l, t_l+\Delta t_l) = f_{j}(x,t_l) - (1/\tau)[f_{j}(x,t_l) - f_{j}^{eq}(x,t_l)]+F_{j}(x,t_l) \Delta t_l
\end{equation}
The corresponding discrete force distribution function can be given by the following relation \cite{guo2002}:
\begin{equation}
F_{j}(x,t_l)=\left(1-(1/2\tau)\right)w_{j}\left[3\Big\{(c_{j}-\mathbf{u_{l}}(x,t_l))/e^{2}\Big\}+9\Big\{(c_{j}\mathbf{u_{l}})/e^{4}\Big\}c_{j}\right].\mathbf{F}(x,t_l)
\end{equation}
where $\mathbf{F}$ is the external force on the liquid phase.

The numerical technqiue to solve the LBE equation with an external force term is as follows \cite{kang11}:\\ \\
First-forcing step: 
\begin{equation}
 \rho_l(x,t_l)\mathbf{u_{l}}(x,t_l)= \sum_{j=1}^{9}c_{j}f_{j}(x,t_l)+(\Delta t_l/2)\mathbf{F}(x,t_l)
\end{equation}
Collision step:
\begin{equation}
f_{j}^{'}(x,t_l)=f_{j}(x,t_l)-(1/\tau)[f_{j}(x,t_l)-f_{j}^{eq}(x,t_l)]
\end{equation}
Second-forcing step:
\begin{equation}
f_{j}^{''}(x,t_l) = f_{j}^{'}(x,t_l)+\Delta t_l F_{j}(x,t_l)
\end{equation}
Streaming step:
\begin{equation}
f_{j}(x+c_{j}\Delta t_l, t_l+\Delta t_l) = f_{j}^{''}(x,t_l)
\end{equation}
\section{Numerical Modelling}
\subsection{Geometry and simulation parameters}
In this study, a three-dimensional (3D) rectangular bubble column  with dimension 0.15 m $\times$ 0.15 m $\times$ 1 m is considered for the simulation. The water is filled up to the height of 0.45 m. Initially, a spherical air-bubble of diameter 4 mm is released in water. It was assumed that the bubble remains spherical throughout the simulation \cite{sung11}.

The aspect ratio ($a_r$) between the cell size $\Delta x$ and bubble diameter $d_b$ was chosen to be 1.25 as given by \cite{niv08}: 
\begin{equation}
a_r = \Delta x /d_b = 1.25
\end{equation} 
The conversion of the physical unit to lattice unit is shown in Appendix A. Simulation conditions for the rising of a single bubble in a stagnant liquid are given in Table 2.
\begin{table}[!h]
\caption{Simulation conditions for bubble of $4$ mm diameter rising in a quiescent water tank }
\centering
\begin{adjustbox}{width=0.6\textwidth}
\small
{\begin{tabular}[l]{@{}llll}
\noalign{\smallskip}\hline\noalign{\smallskip}
Physical domain & & & $0.15$ m $\times$ $0.15$ m $\times$ $0.45$ m\\
Computational domain & & & $30$ $\times$ $30$  $\times$ $90$ \\
Liquid time step & & &$0.0001 $ sec\\
Bubble time step & & &$0.00001$ sec\\
Simulation time & & &$2.0000$ sec\\
\noalign{\smallskip}\hline
\end{tabular}}
\label{symbols}
\end{adjustbox}
\end{table}
The physical properties for the continuous (water) and dispersed (air-bubble) phase used in this simulation are given in Table 3.
\begin{table}[!h]
\caption{Physical properties of dispersed air bubble and continuous liquid phase \cite{deen04}}
\centering
\begin{adjustbox}{width=1\textwidth}
\small
{\begin{tabular}[l]{@{}l l l l l l}
\hline\noalign{\smallskip}
 \textbf{Phase} & \textbf{Property} & & \textbf{Unit} & & \textbf{Value}\\
\noalign{\smallskip}\hline\noalign{\smallskip}
 Dispersed phase (air-bubble) & Density ($\rho_b$)   & &$kg/m^{3}$ & &$1.0$\\
                             & Diameter ($d_b$) & & $m$& &$0.004$\\
                             & Viscosity ($\mu_b$) & & $kg/m-s$ & &$1.8$ $\times$ $10^{-5}$\\ 
Continuous phase (water)     & Density ($\rho_l$)  & &$kg/m^{3}$ & &$1000$\\
                             & Viscosity ($\mu_l$) & &$kg/m-s$ & &$0.001$\\ 
                             & Surface tension ($\sigma$) & &$N/m$ & &$0.073$\\
\noalign{\smallskip}\hline
\end{tabular}}
\label{symbols}
\end{adjustbox}
\end{table}
\subsection{Interphase coupling}
A two-way coupling between the continuous and dispersed phase is done by the ”cheap clipped fourth-order polynomial mapping function”. The mapping function was introduced by \cite{deen04}, which translates the influence of Eulerian quantities on Lagrangian position and vice-versa. The mapping function should satisfy the following criteria given below \cite{kit01}:
\begin{itemize}
  \item It should be a smooth function, i.e. the first derivation should be continuous.
  \item It should have an absolute maximum around the position where the variable is transferred.
  \item For practical reasons, it should have a finite domain. At the boundaries of the domain, the function should be zero.
  \item The integral of the function over the entire domain should equal unity.
\end{itemize}
The mapping function for this two-phase coupling is given as \cite{deen04, darm06}:
\begin{equation}
Z(x_l-x_b) = \begin{cases}
\frac{15}{16}[\frac{(x_l-x_b)}{n^5} - 2\frac{(x_l-x_b)}{n^3} + \frac{1}{n}, & -n\leq((x_l-x_b))\leq n\\
0, & otherwise
\end{cases}
\end{equation}
where $x_b$ is the position vector of gas bubble and $n$ = 1.5$d_b$ is the width of the mapping window. For the three-dimensional (3D) domain, the influence of eulerian quantities (i.e. \textit{velocity}, \textit{vorticity}) on bubble position is evaluated using the given relation \cite{deen04}:
\begin{equation}
\int_{\Omega_j}Z{d}{\Omega} =\int_{\Omega_{j,y}}\int_{\Omega_{j,x}}Z(x_l-x_b)Z(y_l-y_b)Z(z_l-z_b)dxdydz
\end{equation}
\begin{equation}
\Psi = \sum_{j}\sigma(j)\int_{\Omega_j}Z{d}{\Omega}
\end{equation}
where,
\begin{itemize}
\item $\Psi$ defines the influenced eulerian quantites at bubble position
\item $x_l$ , $y_l$, $z_l$ are the position cordinates of liquid computational cell $j$ \item $x_b$ , $y_b$, $z_b$ are the bubble centroids
\item $\sigma$ be the corresponding eulerian quantity 
\end{itemize}
Similarly, the influence of lagrangian quantity on the liquid computational cell $j$ is calculated using the formula:
\begin{equation}
\Phi(j)= \psi_b\int_{\Omega_j}Z{d}{\Omega}
\end{equation}
where, $\psi_b$ is the reaction of the momentum transfer exerted on the bubble, i.e. $\psi_b$ = -$\sum \mathbf{F}$

\begin{algorithm}[H]
 \caption{LBGK-LPT approach to model dispersed gas-liquid flows}
 \begin{algorithmic}[1]
 \renewcommand{\algorithmicrequire}{\textbf{Input:}}
 \renewcommand{\algorithmicensure}{\textbf{Output:}}
 \REQUIRE Physical properties of continuous (liquid) and dispersed (gas) phase
 \ENSURE  out
 \\ \textit{Initialisation} : Calculating PDDF in velocity space\\
 $f_{j} = w_{j}\rho_l[1+\frac{3}{e^2}(c_{j}.\mathbf{u_l})+\frac{9}{2e^4}(c_{j}.\mathbf{u_l})^2-\frac{3}{2e^2}\mathbf{u_l}^2$  
 \\ \textit{LOOP Process}
  \FOR {$i = 1$ to $t$}
  \STATE Tracked bubble position in the domain and select the nearest eulerian node which interact with bubble.
  \STATE \textbf{Forward coupling:} mapping of eulerian (liquid) quantities on lagrangian (air-bubble) position, i.e. \textit{velocity}, \textit{vorticity}:
  $\Psi = \sum_{j}\sigma(j)\int_{\Omega_j}Z {d}{\Omega}  $
  \STATE \textbf{Net force on bubble:} calculate total force acting on bubble i.e. \textit{buoyancy}, \textit{stress gradient}, \textit{drag}, \textit{lift} and \textit{virtual mass} using: \\
  $\mathbf{F_b} = \mathbf{F_G }+ \mathbf{F_S}+\mathbf{F_D} + \mathbf{F_L} + \mathbf{F_{VM}}$ 
  \STATE \textbf{Update bubble velocity and postion:} The velocity and position of bubble can be updated using:\\
  $\mathbf{u_b}^{n+1} = \mathbf{u_b}^n + \frac{\mathbf{F_b}}{m_b}\Delta t_b$, $x_b^{n+1} = x_b^{n} + S$
  \STATE \textbf{Backward coupling:} Mapped the reaction force calulated from the updated lagrangian (i.e. bubble) velocity on the eulerian (liquid) cell using:\\
  $\psi_{b} =  -(\mathbf{F_D} + \mathbf{F_L}$) = $-\mathbf{F}$\\
  $\Phi(j)= \psi_b\int_{\Omega_j}Z{d}{\Omega}$ 
    \STATE \textbf{Discrete force distribution function:} The reaction force that mapped on eulerian node from lagrangian frame is defined in the discrete form as:\\
    $ F_{j}(x,t) = (1-\frac{1}{2\tau})w_{j}[3\frac{c_{j}-\mathbf{u_l}(x,t)}{e^2}+9\frac{c_{j}\mathbf{u_l}(x,t)}{e^4}c_{j}]*\Phi(j) $\\
    \STATE \textbf{Equilibrium, PDDF:} The equilibrium, PDDF can be calculated as:\\
   $f_j^{eq} = w_{j}\rho_l[1+\frac{3}{e^2}(c_{j}.\mathbf{u_l})+\frac{9}{2e^4}(c_{j}.\mathbf{u_l})^2-\frac{3}{2e^2}\mathbf{u_l}^2]$
   \STATE perform first forcing step, collision step, second forcing step and streaming step on eulerian computational cell  given in section $\mathit{2.2}$, provide boundary-conditions and update macroscopic properties for liquid i.e. $velocity$, $momentum$  $density$   
  \IF {bubble reaches top layer of fluid}
  \STATE stop
  \ENDIF
  \ENDFOR
 \RETURN $2$
 \end{algorithmic}
 \end{algorithm}
\subsection{Boundary-conditions}
A no-slip boundary condition based on non-equilibrium half-way bounce-back condition given by \cite{zou97} are applied to every side of the computational domain except for the top boundary of the domain where a free-slip boundary condition is applied \cite{sung11}. Algorithm 1 explains the detailed procedure for simulating the dispersed gas-liquid flows using LBGK-LPT model.

\section{Results and Discussion}
\subsection{Validation test}

In this section, simulation results of a single bubble rising in a 3D rectangular liquid column using LBGK-LPT approach are compared with theoretical and available experimental data in the literature. The relative velocity of bubble motion at different time instant can be calculated theroretically using the expression below:
\begin{equation}
d\mathbf{v_{r}}/dt = (\mathbf{F_B}+\mathbf{F_D})/m_b
\end{equation} 
where $\mathbf{v_r}$ be the velocity of bubble motion relative to the velocity of the liquid phase.

The terminal velocity of the bubble is calculated theoretically using the force balance equation. A bubble rising in a quiescent liquid attains a constant velocity when the net gravity force will be equal to the drag force on the bubble. The force balance equation can be defined as:
\begin{equation}
(\rho_{l}-\rho_{b})V_{b}g = (1/2)*(C_{D}v_{T}^{2}\pi r_{b}^{2})
\end{equation}
where, $v_{T}$ represent the terminal velocity of bubble. Solving Eq.(23) gives an expression for terminal velocity:
\begin{equation}
v_{T} = \sqrt{\Big\{(\rho_{l}-\rho_{b})V_{b}g\Big\}/\Big\{(1/2)(C_{D}\pi r_{b}^{2})\Big\}}
\end{equation}
The results are also validated for terminal velocity at a different case of bubble diameter with Mendelson equation given by \cite{mend67}. The equation was given as \cite{krish99}:
\begin{equation}
v_{T_M} = \sqrt{\Big\{2\sigma/(\rho_{l}d_{b})\Big\}+\Big\{gd_{b}/2\Big\}}
\end{equation}
\begin{table}[!h]
\caption{Comparison of predicted, theoretical and experimental terminal velocity at a different case of bubble diameter \cite{krish99}, \cite{cli78}}
\centering
\begin{adjustbox}{width=1\textwidth}
\small
{\begin{tabular}[l]{@{}c c c c c c c c}
\hline\noalign{\smallskip}
 \thead{\textbf{Bubble }\\ \textbf{diameter} \\ \textbf{(mm)}} & \thead{\textbf{Simulation} \\ \textbf{(LBGK-LPT)}} & \thead{\textbf{Experimental}\\ \textbf{(Clift.et.al)}} & \thead{\textbf{Force}\\ \textbf{balance} \\ \textbf{eq.}}&  \thead{\textbf{Mendelson}\\ \textbf{eq.}} & \thead{\textbf{Error \%}\\ \textbf{with}\\ \textbf{(experimental}\\ \textbf{data)}}&\thead{\textbf{Error \%}\\ \textbf{with}\\ \textbf{(force}\\ \textbf {balance eq.)}} &  \thead{\textbf{Error \%}\\ \textbf{with}\\ \textbf{(Mendelson}\\ \textbf{ eq.)}}\\
\noalign{\smallskip}\hline\noalign{\smallskip}
2  & 0.2898 & 0.3000 & 0.2877 & 0.2878 & 3.40 & 0.73 & 0.69\\
3  & 0.2533 & 0.2652 & 0.2517 & 0.2518 & 4.49 & 0.63 & 0.59\\
4  & 0.2383 & 0.2509 & 0.2368 & 0.2369 & 5.02 & 0.63 & 0.59\\
5  & 0.2331 & 0.2416 & 0.2317 & 0.2318 & 3.52 & 0.60 & 0.56\\
6  & 0.2332 & 0.2355 & 0.2318 & 0.2319 & 0.97 & 0.60 & 0.56\\
7  & 0.2363 & 0.2386 & 0.2348 & 0.2349 & 0.96 & 0.64 & 0.59\\
8  & 0.2411 & 0.2396 & 0.2397 & 0.2398 & 0.63 & 0.58 & 0.54\\
9  & 0.2471 & 0.2447 & 0.2456 & 0.2457 & 0.98 & 0.61 & 0.57\\
10 & 0.2538 & 0.2447 & 0.2521 & 0.2523 & 3.71 & 0.67 & 0.59\\
\noalign{\smallskip}\hline
\end{tabular}}
\label{symbols}
\end{adjustbox}
\end{table}
The comparison of simulated, theoretical and experimental terminal velocity at a different case of bubble diameter shown in Table 4. The experimental data were extracted for the case of bubble rising in pure water from \cite{cli78} using the g3data software. 
\begin{figure}[H]
\begin{center}
\begin{adjustbox}{width=1\textwidth}
\small
\begin{minipage}{\textwidth}
\subfigure[]{
{\includegraphics[width=0.5\textwidth]{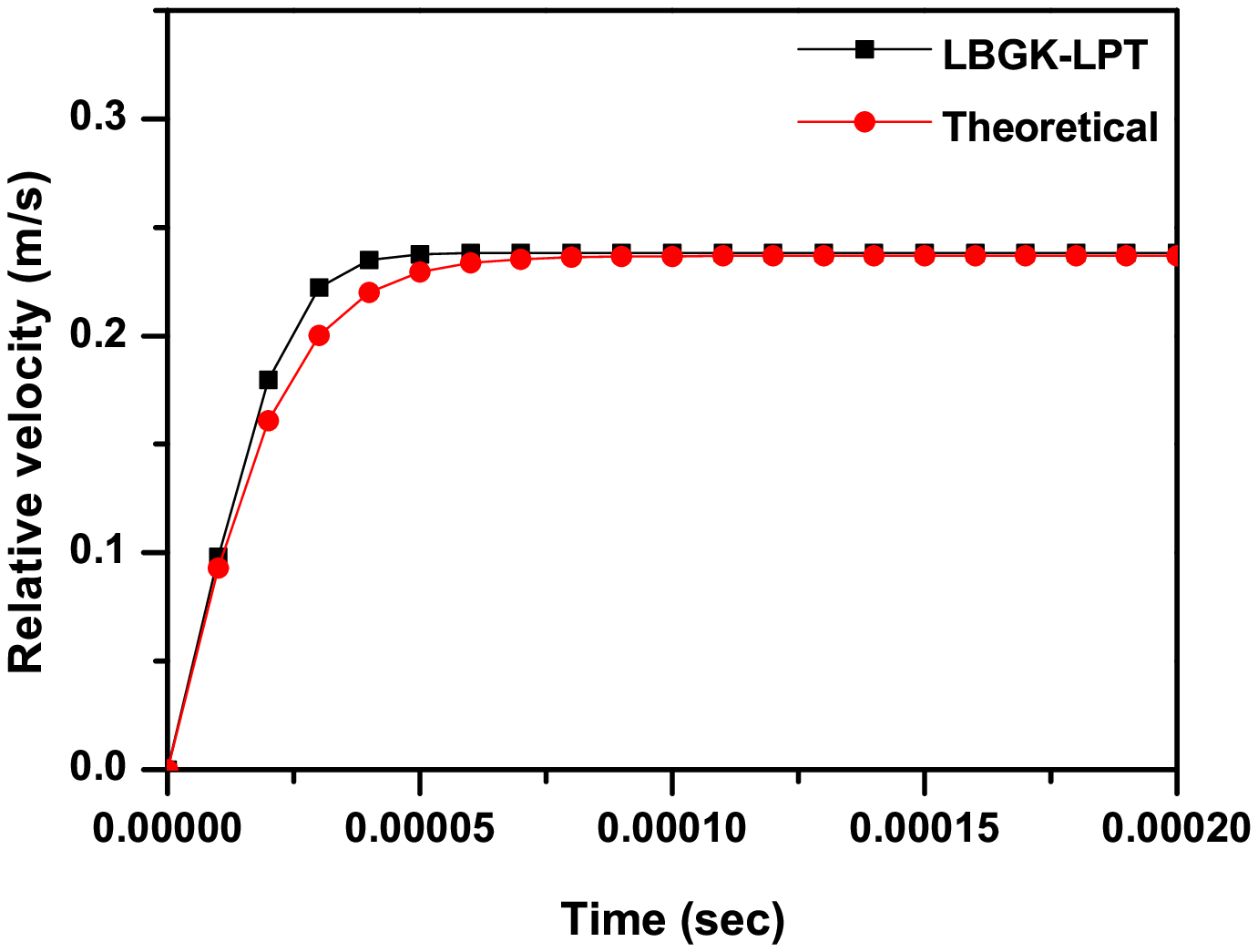}}}\hspace{0.2pt}
\subfigure[]{
{\includegraphics[width=0.5\textwidth]{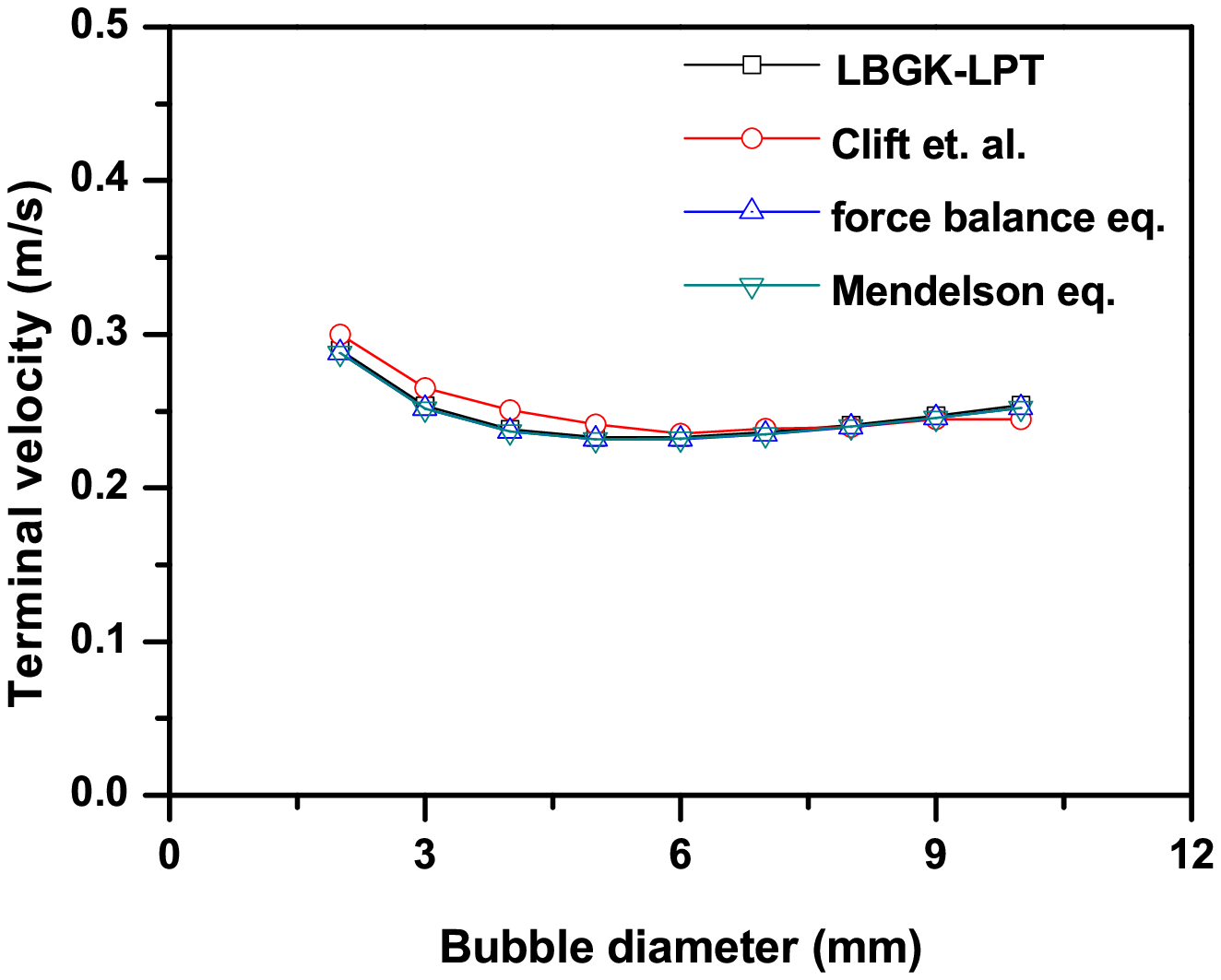}}}
\caption{Comparison of (a) simulated and theoretical relative velocity of 4
mm bubble diameter at different time instant rising in stagnant liquid, (b)
simulated, theoretical and experimental teriminal velocity at a different
case of bubble diameter}
\label{sample-figure}
\end{minipage}
\end{adjustbox}
\end{center}
\label{fig:2}
\end{figure}
Fig.2(a) shows the comparison of simulated and theoretically calculated relative velocity for 4 mm air bubble rising in a quiescent water tank as a function of time. It is seen that after 0.00005 seconds bubble rising with a constant relative velocity. The results obtained from simulation are satisfactory with the theoretical results. Also, it has been observed that bubble takes $\approx$ 1.86 seconds to reach the top layer of fluid.

Fig.2(b) shows the comparison of simulated terminal velocity from the LBGK-LPT model with the corresponding theoretically and experimentally computed terminal velocity of a bubble motion in a stagnant water tank at varying bubble diameter. It is observed that the simulated results are in good agreement with the theoretical results computed from the eq.(23), (24) and with the experimental results of \cite{cli78}. 

Fig.3 shows the velocity fluctuation in continuous liquid phase relative to the motion of air bubble at a different time instant. It is seen that that vortex appeared in the wake region is closer to the edge of a dispersed air bubble.
\begin{figure}[H]
\begin{center}
\begin{adjustbox}{width=1\textwidth}
\small
\begin{minipage}{\textwidth}
\subfigure[]{
{\includegraphics[width=0.5\textwidth]{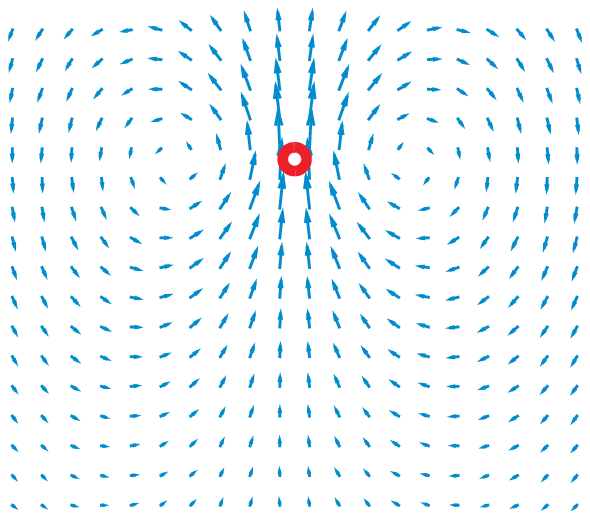}}}\hspace{0.2pt}
\subfigure[]{
{\includegraphics[width=0.5\textwidth]{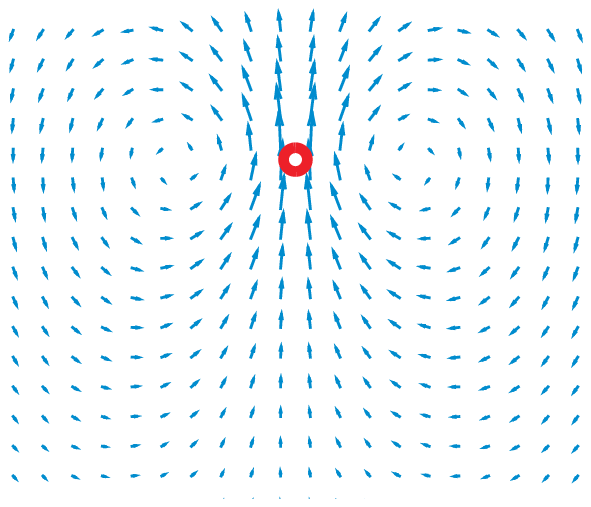}}}\hspace{0.2pt}
\subfigure[]{
{\includegraphics[width=0.5\textwidth]{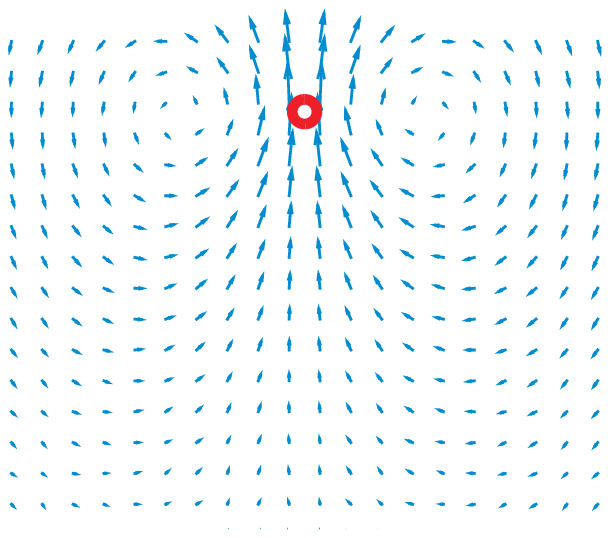}}}\hspace{0.2pt}
\subfigure[]{
{\includegraphics[width=0.5\textwidth]{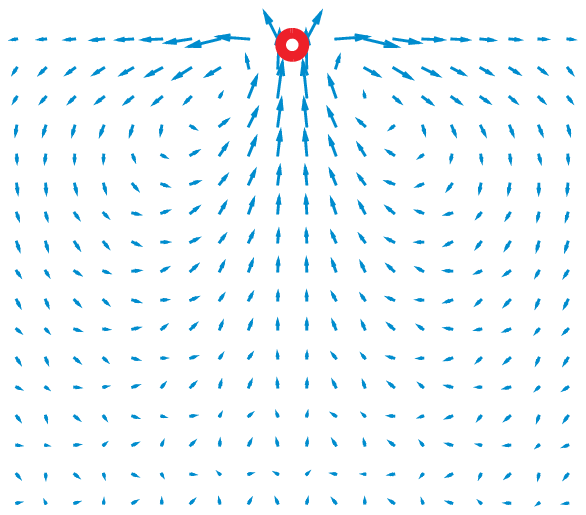}}}
\caption{\label{fig2} 2D representation of liquid velocity fluctuation on rising of air bubble at time (a) t = 0.4632 sec, (b) t = 0.9264 sec, (c) t = 1.3896 sec, and (d) t = 1.85280 sec }
\label{sample-figure}
\end{minipage}
\end{adjustbox}
\end{center}
\label{fig:3}
\end{figure}

\subsection{Grid independence test}
As grid size depends on aspect ratio (from eq. (17)), the different aspect ratio was chosen to investigate grid independence test for different cases of bubble diameter. Table 5, shows the grid size at varying aspect ratio for three different cases of bubble diameter (i.e. 4 mm, 8 mm, and 10 mm) and the predicted terminal velocity from simulation for these cases are shown in Table 6. 

\begin{table}[!h]
\caption{Unit converison parameters at different range of aspect ratio}
\centering
\begin{adjustbox}{width=1\textwidth}
\small
{\begin{tabular}[l]{@{}l l l l l c c c c c}
\hline\noalign{\smallskip}
\thead{\textbf{Aspect}\\ \textbf{ratio}} & \thead{\textbf{LCF}} & \thead{\textbf{MCF}} & \thead{\textbf{TCF}} & \thead{\textbf{cell size} \\ \textbf{($\Delta x$)}}&  \thead{\textbf{column width}\\ \textbf{in}\\ \textbf{ lattice unit}\\$w_{lu}$} &
\thead{\textbf{column depth}\\ \textbf{in}\\ \textbf{ lattice unit}\\ $d_{lu}$} & \thead{\textbf{water level}\\ \textbf{in}\\ \textbf{ column tank}\\ $h_{lu}$} & \thead{\textbf{bubble diameter} \\ \textbf{(physical unit)} \\ $d_{b,pu}$ (mm)}& \thead{\textbf{bubble}\\ \textbf{diameter} \\ \textbf{(lattice unit)} \\ $d_{b,lu}$}\\
\noalign{\smallskip}\hline\noalign{\smallskip}
$1.25$      & 200    & 8000      & 10000    & 0.005000    & 30  & 30 & 90  & 4  & 0.8\\
$0.625$     & 400    & 64000     & 10000    & 0.002500 = $\Delta x/2$    & 60 & 60 & 180 & 4  & 1.6\\
$0.3125$    & 800    & 512000    & 10000    & 0.001250 = $\Delta x/4$    & 120 & 120 & 360 & 4  & 3.2\\
$1.25$      & 100    & 1000      & 10000    & 0.010000    & 15  & 15 & 45 & 8  & 0.8\\
$0.625$     & 200    & 8000      & 10000    & 0.005000 = $\Delta x/2$    & 30  & 30 & 90 & 8  & 1.6\\
$0.3125$    & 400    & 640000    & 10000    & 0.002500 = $\Delta x/4$   & 60 & 60 & 180 & 8  & 3.2\\
$1.25$      & 80     & 512       & 10000    & 0.012500    & 12  & 12 & 36 & 10 & 0.8\\
$0.625$     & 160    & 4096      & 10000    & 0.006250 = $\Delta x/2$    & 24  & 24 & 72 & 10 & 1.6\\
$0.3125$    & 320    & 32768     & 10000    & 0.003125 = $\Delta x/4$    & 48 & 48 & 144 & 10 & 3.2\\
\noalign{\smallskip}\hline
\end{tabular}}
\label{symbols}
\end{adjustbox}
\end{table}
\begin{figure}[H]
\centering
\begin{adjustbox}{width=1\textwidth}
\small
\subfigure[]{
\includegraphics[width=1\textwidth]{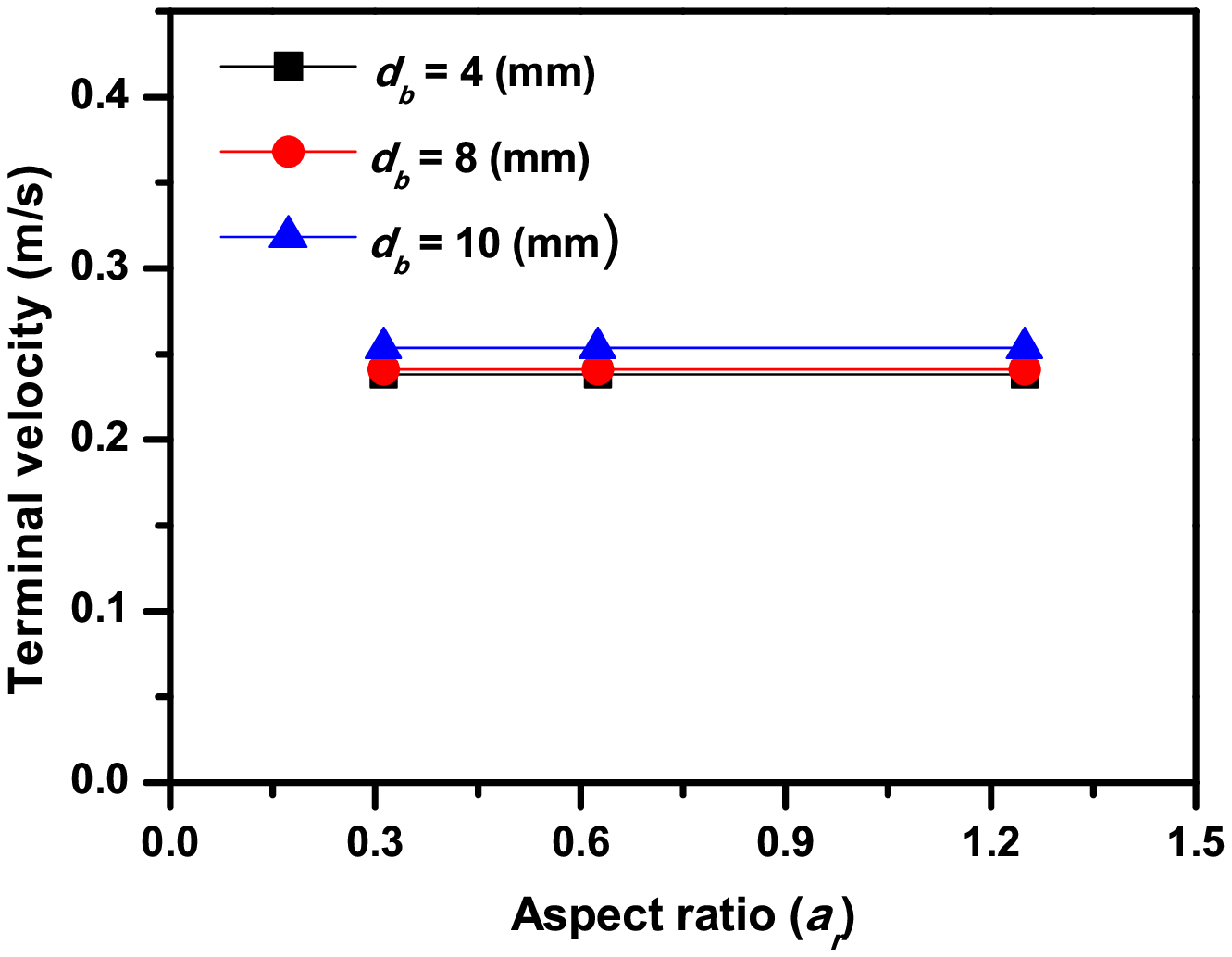}}\
\subfigure[]{
\includegraphics[width=1\textwidth]{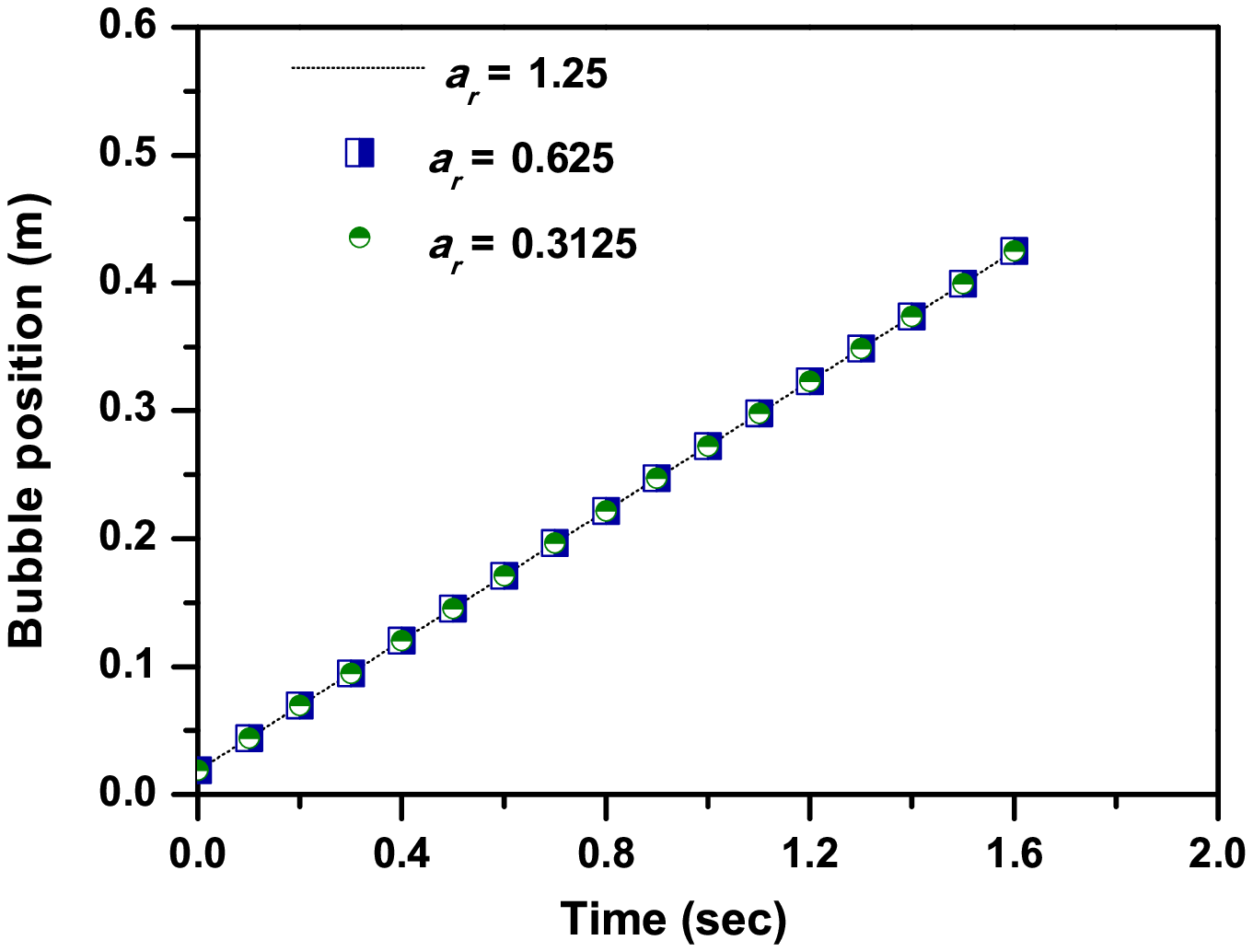}}\
\subfigure[]{
\includegraphics[width=1\textwidth]{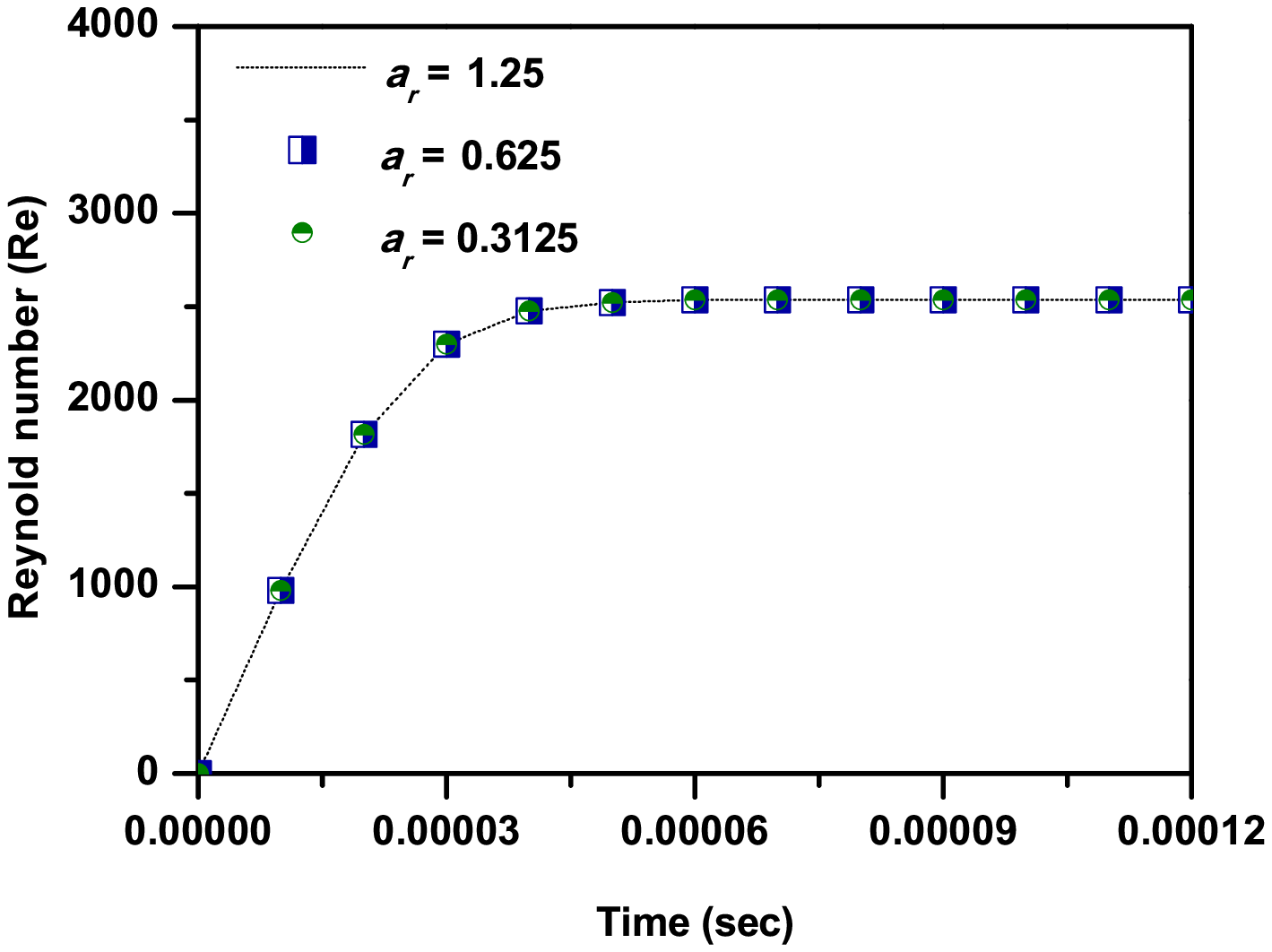}}
\end{adjustbox}
\caption{Grid independence test (a) terminal velocity at varying aspect ratio for three different cases of bubble diameter, (b) bubble position vs time at different aspect ratio for 10 mm bubble diameter, and (c) reynold number (Re) vs time at different aspect ratio for 10 mm bubble diameter}
\label{fig:4}
\end{figure}
\begin{table}[!h]
\caption{Terminal velocity at varying aspect ratio for three different cases of bubble diameter }
\centering
\begin{adjustbox}{width=1\textwidth}
\small
{\begin{tabular}[l]{@{}c c c c c c}
\hline\noalign{\smallskip}
\thead{\textbf{Aspect ratio}} & \thead{\textbf{Grid size}} & \thead{\textbf{bubble diameter} \\ \textbf{(physical unit)} \\ $d_{b,pu}$ (mm)} & \thead{\textbf{bubble diameter} \\ \textbf{(lattice unit)} \\ $d_{b,lu}$} & \thead{\textbf{Terminal} \\ \textbf{velocity}}\\ 
\noalign{\smallskip}\hline\noalign{\smallskip}
$1.2500$    &     30   $\times$ 30  $\times$ 90    & 4       & 0.8  & 0.2383\\
$0.6250$    &     60   $\times$ 60  $\times$ 180   & 4       & 1.6  & 0.2383\\
$0.3125$    &     120  $\times$ 120 $\times$ 360   & 4       & 3.2  & 0.2383\\
$1.2500$    &     15   $\times$ 15  $\times$ 45    & 8       & 0.8  & 0.2411\\
$0.6250$    &     30   $\times$ 30  $\times$ 90    & 8       & 1.6  & 0.2411\\
$0.3125$    &     60   $\times$ 60  $\times$ 180   & 8       & 3.2  & 0.2411\\
$1.2500$    &     12   $\times$ 12  $\times$ 36    & 10      & 0.8  & 0.2538\\
$0.6250$    &     24   $\times$ 24  $\times$ 72    & 10      & 1.6  & 0.2538\\
$0.3125$    &     48   $\times$ 48  $\times$ 144   & 10      & 3.2  & 0.2538\\
\noalign{\smallskip}\hline
\end{tabular}}
\label{symbols}
\end{adjustbox}
\end{table}

Fig.4(a) shows the terminal velocity of a bubble at varying aspect ratio for three different cases of bubble diameter (i.e. 4 mm, 8 mm and 10 mm). The results within the grid size 30 $\times$ 30 $\times$ 90 from aspect ratio 1.25 are same as that of the grid size 60 $\times$ 60 $\times$ 180 and 1200 $\times$ 120 $\times$ 360 from aspect ratio 0.625 and 0.3125 respectively, for the case of 4 mm bubble diameter. For other two cases of 8 mm and 10 mm bubble diameter, results are also similar for different grid sizes. For the case of 10 mm bubble diameter: the position of the bubble and Reynolds number (Re) on a particular instant of time at different aspect ratio was evaluated and results shown in Fig.4(b) and 4(c) also demonstrate model independence from the grid size.
\subsection{Stability test}
The different range of dimensionless relaxation frequency ($\omega$) of LBGK model is applied for investigating the maximum value up to which the solution is stable.  For the current study of dispersed gas-liquid flow, the predicted maximum allowed collision frequency for a stable solution is shown in Fig.5. Fig.5 shows the velocity of bubble motion relative to liquid phase velocity as a function of time for a given range of relaxation frequency. As can be seen from Fig.5, the solution goes unstable for a relaxation frequency of 1.98, means the maximum allowed value of relaxation frequency for a stable solution of dispersed gas-liquid flow is 1.97. This depicts that solution becomes unstable for a very low range of liquid phase viscosity (from relation in eq. (9)).

\begin{figure}
\begin{center}
\includegraphics[width=1\textwidth]{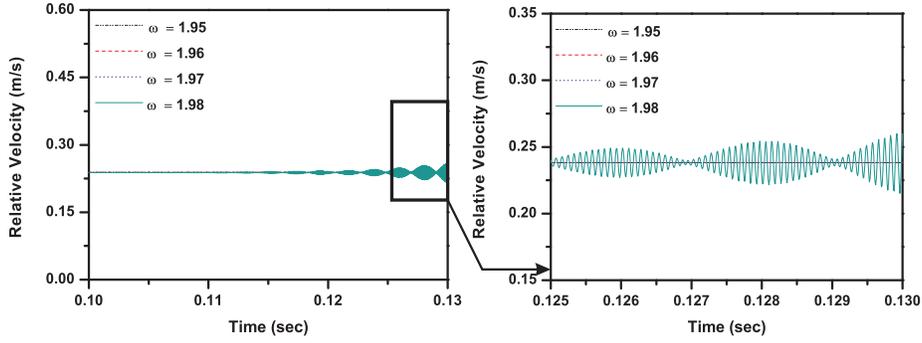} 
\end{center}
\label{fig:5}
\caption{Numerical stability test for dispersed gas-liquid flow using LBGK-LPT model in the parameter space of bubble relative velocity vs time and at a given value of $\omega$}
\end{figure}
\section{Concluding Remarks}
A modeling technique to simulate the benchmark problem of dispersed gas-liquid flow (i.e. single bubble rising in a stagnant liquid) using EL approach has been presented. The BGK scheme of lattice Boltzmann method (LBM) proposed by \cite{bhat54} was used to discretize the continuous liquid phase. The dispersed gas (air-bubble) phase modeled with the lagrangian particle tracking (LPT)  approach. The concept of two-way coupling using cheap-clipped polynomial mapping function given by \cite{deen04} was used for momentum-transfer between two phases. 

We have demonstrated the numerical accuracy of the model by comparing the simulation results for a relative and terminal velocity of the bubble with theoretical and experimental results. The results are in good agreement with the theoretical and available experimental data. Additionally, the velocity fluctuation in the continuous liquid phase due to bubble motion is also presented at a different time instant. The results obtained from the grid independence study shows that model is independent of the grid size. We also performed stability test for the model, and result shows that the model goes unstable for low value of liquid phase viscosity. 

Further, the bubble tracking module validated in current work will be used to develop a solver capable of simulating an industrial scale bioreactor with all the complex physical process.

\section*{References}

\bibliography{mybibfile}
\appendix
\section{Physical to lattice unit conversion}
The simulation of continuous liquid phase through lattice Boltzmann method (LBM) requires the conversion of physical unit to lattice unit. This can be performed as:
\begin{equation}
C.F.= lu/pu
\end{equation}
where $C.F.$ is the conversion factor for the conversion of physical unit ($pu$) to lattice unit ($lu$). With the reference of eq. (A.1), the corresponding length ($L.C.F$), time ($T.C.F$) and mass ($M.C.F.$) conversion factor can be defined as:
\begin{equation}
L.C.F. = l_{lu}/l_{pu}, \ \ \ 
T.C.F. = t_{lu}/t_{pu},\ \ \
M.C.F. = m_{lu}/m_{pu} 
\end{equation}
From the geometrical parameters, we have
\begin{equation}
h_{pu} = 0.45 \ mm, w_{pu} = 0.15 \ mm, d_{pu} = 0.15 \ mm
\end{equation}
where $h_{pu}$, $w_{pu}$ and $d_{pu}$ be the level of water in rectangular column tank, width and depth column tank in physical unit respectively. According to \cite{niv08},  the aspect ratio (i.e. the ratio between the grid spacing of continuous liquid domain to bubble diameter) was chosen to be 1.25. Thus,
\begin{equation}
\Delta x_{pu}/d_{b,pu} = 1.25
\end{equation}
where $\Delta x_{pu}$ and $d_b$ represents the grid spacing and bubble diameter respectively. Solving eq. (A.4) for the case of 4 mm bubble diameter, we can obtain:
\begin{equation}
\Delta x_{pu} = 0.005
\end{equation}
From eq. (B.2), we have:
\begin{equation}
L.C.F.= \Delta x_{lu} / \Delta x_{pu}
\end{equation}
For the lattice boltzmann method, the lattice size ($\Delta x_{lu}$) and the time step size ($\Delta t_{lu}$) are
\begin{equation}
\Delta x_{lu} = \Delta t_{lu} = 1
\end{equation}
Thus, using the value of $\Delta x_{pu}$ from eq. (A.5),  value of $L.C.F$ can be easily evaluated from eq. (A.7) gives:
\begin{equation}
L.C.F. = 200
\end{equation}
From eq. (A.2), (A.3), and (A.8) we have
\begin{equation}
h_{lu} = 90, w_{lu} = 30, d_{lu} = 30, d_{b,lu} = 0.8
\end{equation}
Where $h_{lu}$ is level of water in column tank, $w_{lu}$ is width of column tank and $d_{b,lu}$ is the diameter of bubble in lattice unit. The liquid time step has taken to be 0.0001 sec in the physical unit for the simulation i.e. $\Delta t_{pu}$. Thus, we compute the $T.C.F$ using the relation from eq. (A.2) we have
\begin{equation}
T.C.F  = \Delta t_{lu}/\Delta t_{pu} = 1/0.0001 = 10000
\end{equation}
The density of continuous (liquid) phase in the lattice unit $\rho_{lu}$ and physical unit $\rho_{pu}$ are taken to be:
\begin{equation}
\rho_{lu} = 1, \rho_{pu} = 1000
\end{equation}
Correspondingly, the $M.C.F$ can be calculated by the given relation:
\begin{equation}
\rho_{pu} = \rho_{lu}\Big(LCF^3/MCF\Big)
\end{equation}
which gives,
\begin{equation}
M.C.F = 8000
\end{equation}
\listoffigures
\end{document}